\documentclass[pra,twocolumn]{revtex4-1}
\usepackage{graphicx}
\usepackage{amsmath}
\usepackage{amssymb}
\usepackage{textcomp}
\usepackage{color}
\usepackage{transparent}

\begin{document}
\title{Characteristics of unconventional Rb magneto-optical traps}
\author{K. N. \surname{Jarvis}}
\email{k.jarvis14@imperial.ac.uk}
\affiliation{Centre for Cold Matter, Blackett Laboratory, Imperial College London, Prince Consort Road, London SW7 2AZ, UK}
\author{B. E. Sauer}
\affiliation{Centre for Cold Matter, Blackett Laboratory, Imperial College London, Prince Consort Road, London SW7 2AZ, UK}
\author{M. R. Tarbutt}
\affiliation{Centre for Cold Matter, Blackett Laboratory, Imperial College London, Prince Consort Road, London SW7 2AZ, UK}
\begin{abstract}
We study several new magneto-optical trapping configurations in $^{87}$Rb. These unconventional MOTs all use type-II transitions, where the angular momentum of the ground state is greater than or equal to that of the excited state, and they may use either red-detuned or blue-detuned light. We describe the conditions under which each new MOT forms. The various MOTs exhibit an enormous range of lifetimes, temperatures and density distributions. At the detunings where they are maximized, the lifetimes of the various MOTs vary from 0.1 to 15~s. One MOT forms large ring-like structures with no density at the centre. The temperature in the red-detuned MOTs can be three orders of magnitude higher than in the blue-detuned MOTs. We present measurements of the capture velocity of a blue-detuned MOT, and we study how the loss rate due to ultracold collisions depends on laser intensity and detuning.
\end{abstract}
\maketitle
\section{Introduction}
It has been over 30 years since the first demonstration of a magneto-optical trap (MOT)~\cite{Prentiss1987}. Since then, the MOT has led to a diverse range of applications and facilitated the emergence of broad fields of productive research. Normally, atomic MOTs operate on type-I transitions, where the hyperfine angular momentum of the excited state ($F'$) exceeds that of the ground state ($F$), $F'=F+1$. Examples also exist of type-II atomic MOTs \cite{Marcassa1996,Shang1994,Nasyrov2001,Tiwari2008}, which have $F'\leq F$, but these have the unfavourable properties of high temperature and low density, and so have not received much attention. Recently, a number of groups have demonstrated magneto-optical trapping of diatomic molecules \cite{Barry2014a,Truppe2017,Anderegg2017}, which are cooled using type-II transitions \cite{Shuman2009}, and this has generated a renewed interest in understanding the properties of type-II MOTs~\cite{Tarbutt2015,Devlin2016}. This new understanding led us to propose and demonstrate a novel MOT, where blue-detuned light drives type-II transitions \cite{Jarvis2018}. This MOT performed exceptionally well, providing a phase-space density far higher than reported in any other type-II MOT, and comparable with the very best type-I MOTs. It is now understood that an appreciable magneto-optical force can exist for transitions that have $F'=F$ or $F'=F-1$, and that there may be many previously unrealized MOT configurations relying on such transitions. To illustrate, consider the alkali metal atoms. When the hyperfine structure is well resolved, each of the two hyperfine ground states is coupled to an excited-state hyperfine level $F'$. There are three choices of $F'$ from the D2 line and two further choices from the D1 line. Furthermore, when $F'=F$ or $F-1$, a stable MOT may be obtained for either red or blue-detuned light, providing the polarisation is chosen correctly. 

Here, we report a number of new magneto-optical trapping configurations observed using the D2 line in $^{87}$Rb, showing eight new combinations of transitions that produce a stable MOT. We illustrate how significantly the properties of these new traps can differ, both from one another and from the widely-studied type-I MOT. The temperature of the atoms in various MOTs can differ by a factor of 1000, and in some cases atoms form striking ring-like structures. We also extend our characterization of the blue-detuned MOT reported previously~\cite{Jarvis2018} by measuring the capture velocity and loss rate coefficients as a function of laser intensity and detuning. This study helps improve our understanding of a ubiquitous and important technique used throughout the atomic physics community, and one that is emerging as a key tool for cooling and trapping molecules. The new MOT configurations may also be valuable for studying ultracold collisions, especially the influence of hyperfine state and laser detuning. For example, we demonstrate configurations where most of the population resides in the lower hyperfine ground-state, in contrast to the usual MOT, and we characterize blue-detuned MOTs where the effects of near-resonant blue-detuned light on ultracold collisions can be studied.

\section{Experiment}

Our experimental setup consists of a simple vapour-cell magneto-optical trap. Two separate lasers provide the cooling and trapping light needed. One drives transitions from the lower hyperfine ground state, and is called ${\cal L}_1$, while the other drives transitions from the upper hyperfine ground state, and is called ${\cal L}_2$. Their detunings from the excited state $F'$ are written $\Delta_{FF'}$. The intensity at the centre of the MOT is $I_t$ and is divided approximately equally between the 6 MOT beams and again between ${\cal L}_1$ and ${\cal L}_2$. A third laser, ${\cal L}_{\rm ref}$, is locked to the $F=2\to F'=1,3$ saturated absorption crossover resonance of $^{87}$Rb. It is used as a frequency reference and for absorption imaging of the atoms. The frequencies of ${\cal L}_{1,2}$ are stabilized relative to ${\cal L}_{\rm ref}$ using frequency-offset locks. The light from ${\cal L}_1$ and ${\cal L}_2$ is overlapped on a 50:50 beamsplitter cube, whose outputs are used to form two independent sets of MOT beams of opposite handedness. A pair of acousto-optical modulators (AOMs) is used to control whether each of the sets of beams is on or off. At the start of an experiment ${\cal L}_2$ is red-detuned from the transition to $F'=3$, while ${\cal L}_1$ is tuned to either $F'=1$ or $F'=2$, both of which can function as repumping transitions. A type-I MOT then loads from the atomic vapour produced by a dispenser. Once this MOT is loaded, the intensity of both ${\cal L}_1$ and ${\cal L}_2$ are reduced and the detuning of ${\cal L}_2$ increased, which further cools the trapped atoms. Then, the light is quickly switched off using the AOM, the lasers are re-locked to new frequencies, and light with the desired handedness is switched back on. Many of the new MOTs require this pre-cooling stage because the Doppler-cooling forces are too weak to capture atoms directly from the room-temperature vapour.

\begin{figure}[t]
	\centering
	\includegraphics{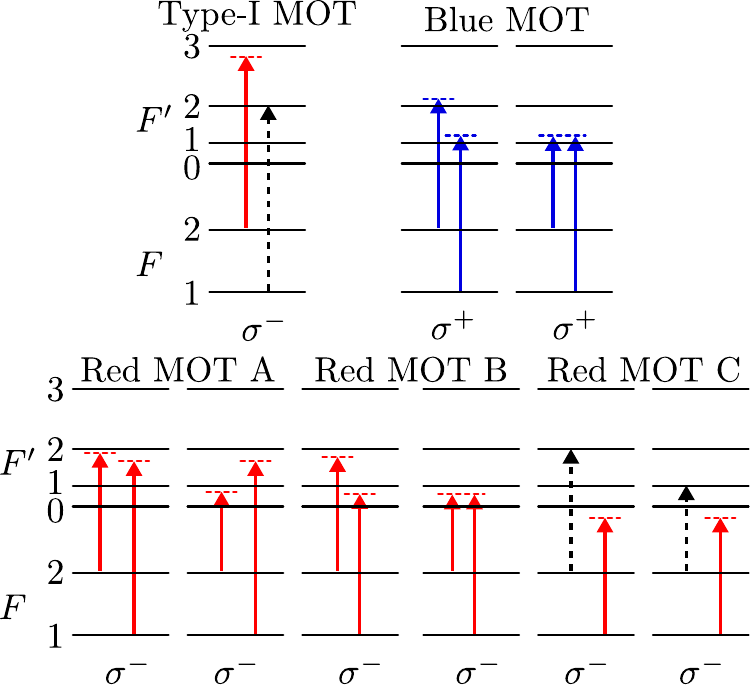}
	\caption{Stable trapping configurations for cold atoms observed in this experiment. We choose the $z$-axis to be along the magnetic field direction. In each case the two lasers have the same polarisation.  In the type-I MOT when $g_{F'}>0$, as in $^{87}$Rb, the laser beams that push displaced atoms back towards the centre should be polarised to drive $\sigma^-$ transitions. This is also true for red MOT A, B and C. For the blue MOT, the polarisations are reversed. When a laser can be tuned to resonance with a transition without adversely affecting the MOT, the arrow representing the laser is coloured black and shown dashed.}
	\label{Fig: TrappingConfigurations}
\end{figure}

Fig.~\ref{Fig: TrappingConfigurations} shows the various configurations for which we have observed a stable MOT. The first trapping scheme illustrated is the type-I MOT. The other configurations are grouped into pairs according to the transition driven by ${\cal L}_1$.  Each pair has one configuration with $F=2\to F'=2$, and one with $F=2\to F'=1$. According to the rules for the choice of polarisation given in \cite{Tarbutt2015}, the position-dependent force for atoms in $F=2$ should change sign between these two configurations. We observe a stable MOT for all the configurations shown, implying that the majority of the confining force is due to transitions from the lower hyperfine ground-state. This observation motivates our categorization of the MOTs.

\section{Blue-detuned MOTs}
We begin by discussing the trapping schemes that use blue-detuned light. The principle of a blue-detuned MOT is explained in detail in \cite{Jarvis2018}. Briefly, when considering a two-level atom or molecule there are two possible combinations of laser detuning and polarisation which produce the same position-dependent force. When the upper-state $g$-factor is positive, and $F'=F$ or $F'=F+1$, the beams that push a displaced atom back towards the centre (which we call the restoring beams) can either be red-detuned and polarised to drive $\Delta m_F=-1$ transitions, or can be blue-detuned and polarised to drive $\Delta m_F=+1$. In type-II systems the Doppler and sub-Doppler velocity-dependent forces have the opposite sign for a given detuning. When the light is red-detuned, there is efficient Doppler cooling, so the capture velocity of the MOT is larger, but polarisation-gradient forces heat the atoms to non-zero equilibrium velocities, so the temperature of atoms in a red-detuned MOT is high. Conversely, with blue-detuning there is efficient sub-Doppler cooling to low temperatures, but atoms moving faster than some critical velocity are heated by the Doppler forces and will evaporate from the MOT, resulting in a low capture velocity. In type-I systems, both Doppler and sub-Doppler cooling require red-detuned light, so no stable MOT exists when the light is blue-detuned.
\begin{figure}[t]
	\centering
	\includegraphics{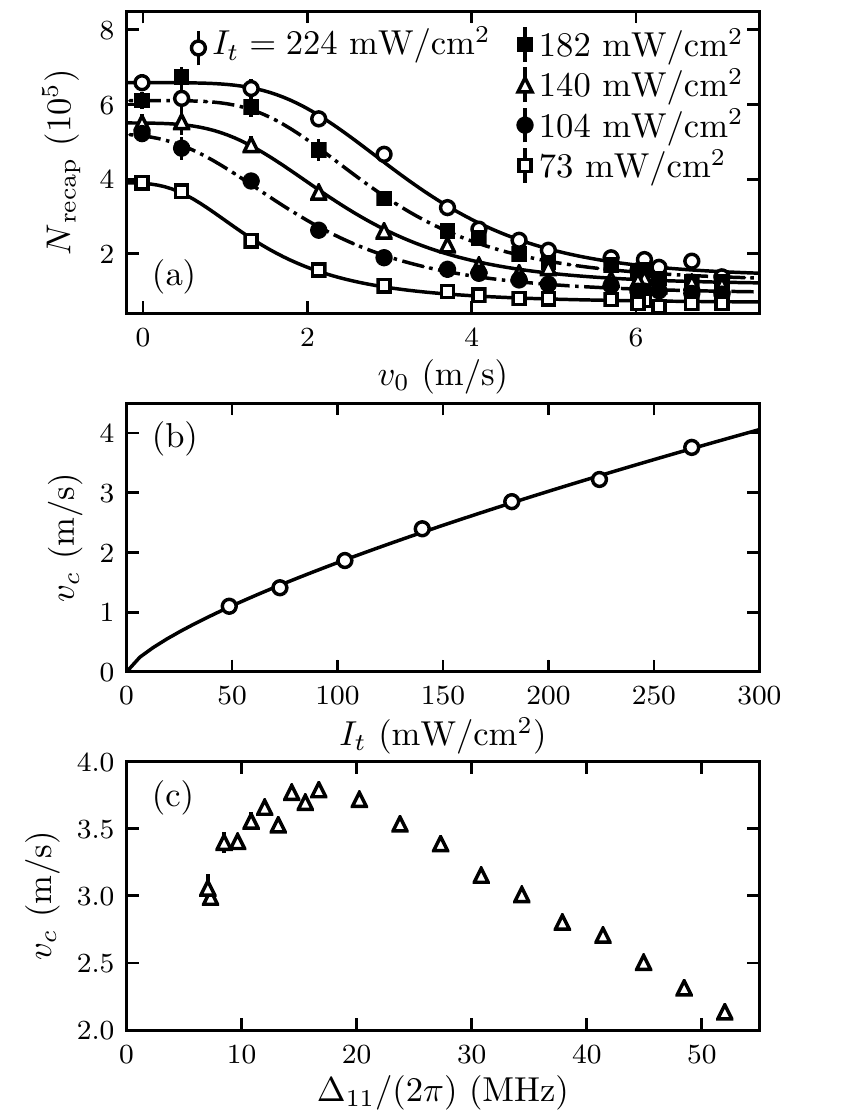}
	\caption{Capture velocity, $v_c$, of the blue-detuned MOT. (a) Number of atoms recaptured by the MOT, $N_{\textrm{recap}}$, as a function of the centre of mass velocity of the cloud, $v_0$. Data are shown for various trap intensities and each dataset is fit to extract the capture velocity at that intensity. (b) Capture velocity as a function of total trap intensity $I_t$. The data are fit to a polynomial $v_c=a I_t^b$, giving a best-fit exponent of $b=0.73(2)$. The other trap parameters are $\Delta_{11}/(2\pi)=+26$~MHz, $\Delta_{22}/(2\pi)=+11.5$~MHz, and $B'=39$~G/cm. (c) Capture velocity as a function of $\Delta_{11}$. The other parameters are $I_t=260$~mW/cm$^2$ and $B'=39$~G/cm.}
	\label{Fig: CaptureVelocity}
\end{figure}

We have observed two configurations of blue-detuned MOT. In the first configuration, presented in \cite{Jarvis2018}, ${\cal L}_1$ is detuned from $F'=1$, and ${\cal L}_2$  is detuned from $F'=2$. In the second combination, both lasers are detuned from $F'=1$. In both configurations the two lasers are polarised so that the restoring beams drive $\Delta m_F=+1$ transitions. This is the opposite polarisation to that used in the type-I MOT. For the MOT to work, ${\cal L}_1$ requires this polarisation according to both the rate-equation model presented in \cite{Tarbutt2015} and the density-matrix model presented in \cite{Devlin2016}. When ${\cal L}_2$ is blue-detuned from $F'=2$, the two approaches again agree that the restoring beam needs to drive $\Delta m_F=+1$ transitions. However, when blue-detuned from $F'=1$, the rate-equation model predicts an anti-confining force for this polarisation, whereas the density-matrix model predicts confinement for the range of magnetic fields explored by an atom close to the trap centre. This difference in the nature of the confining force near the trap centre was highlighted previously~\cite{Devlin2016}. 

The properties of the two blue-detuned MOTs are similar to one another. Both produce clouds of a similar size. With $I_t = 400$~mW/cm$^2$, $\Delta_{11}/(2\pi)=+35$~MHz, and $B'=48$~G/cm the temperature varies between 180~$\mu$K and 215~$\mu$K as $\Delta_{21}/(2\pi)$ is varied between 27 and 37~MHz. At a lower intensity, $I_t = 30$~mW/cm$^2$, the temperature varies between 32 and 38~$\mu$K as $\Delta_{21}/(2\pi)$ is varied between 35 and 47 MHz. The first blue-detuned MOT configuration was characterized in some detail in \cite{Jarvis2018}. Here, we extend that study by measuring the capture velocity and the rate coefficient for trap loss due to ultracold collisions in the MOT.

\subsection{Capture velocity of the blue-detuned MOT}
The capture velocity, $v_c$, is an important parameter of a MOT that governs the steady-state trapped atom number and atom density through the loading rate of the trap \cite{Monroe1990,Gibble1992}. Measurements of the capture velocity can also help to understand which inelastic processes can lead to trap loss \cite{Hoffmann1996}. We measure the capture velocity using the following procedure. First, atoms are cooled and captured in the red-detuned type-I MOT and loaded into the blue-detuned MOT. Then, the MOT light is switched off using an AOM. Next, atoms are optically pumped into $F=2$ using a pulse of light resonant with the $F=1\to F'=2$ transition. Atoms are then launched horizontally through the centre of the MOT by a push beam resonant with the $F=2\to F'=3$ cycling transition. The velocity of the atoms can be tuned by varying the push beam duration, and is calibrated by recording a series of absorption images of the moving atom cloud after the push is complete. From these images the centre of mass velocity of the cloud is extracted. Once the atoms have been launched, the MOT is quickly switched back on and the atoms with $v<v_c$ are recaptured. After a short hold time the number of atoms in the MOT is recorded. This is repeated as the centre-of-mass velocity of the cloud of atoms launched into the MOT is increased, from which the capture velocity of the trap can be inferred. Some data illustrating this is shown in Fig.~\ref{Fig: CaptureVelocity}(a). For each of the sets of data the number of atoms recaptured, $N_{\mathrm{recap}}$, is fit to a function 
\begin{equation}
\begin{aligned}
N_{\textrm{recap}}(v_0) &= N_{0}(F=1) + \\&\frac{N_0(F=2)}{\sqrt{\pi}\sigma_{v_0}}\int_{-v_c}^{v_c}\exp\left[-\frac{(v-v_0)^2}{2\sigma(v_0)^2}\right]dv,
\end{aligned}
\end{equation}
where $N_0(F)$ is the number of atoms in the ground-state hyperfine level labeled by $F$ immediately before the push, $v_0$ is the centre of mass velocity of the atoms in $F=2$ after the push, $\sigma(v_0)$ characterizes the width of the velocity distribution after the atoms are launched, and $v_c$ is the capture velocity of the MOT along the trajectory of the launched atoms and is a free parameter in the fit. The velocity distribution is broadened by the gradient in the intensity of the push beam across the atom cloud, so its width $\sigma$ depends on $v_{0}$ through the push beam duration. This effect visibly skews the roll-off in the recaptured atom number, leading to the high-velocity tails in Fig.~\ref{Fig: CaptureVelocity}(a). To account for this, we find it sufficient to let the width $\sigma$ have a simple linear dependence on $v_0$, $\sigma(v_0)=\sigma_0+a v_0$, where $\sigma_0$ and $a$ are free parameters in the fit. In the measurements presented here atoms are held in the blue-detuned MOT for a total period of 200~ms. Some atoms are lost during this period, and the fraction lost depends on the intensity. This is why $N_{\textrm{recap}}(0)$ depends on intensity. This has no effect on the measurement of $v_c$.

Figure \ref{Fig: CaptureVelocity}(b) shows the capture-velocity of the blue-detuned MOT as a function of $I_t$. The points are extracted from the best-fit parameters obtained from the curves in  part (a). As expected, the capture velocity increases with increasing intensity. Fitting these data to the model $v_c=a I_t^b$ yields $b=0.73(2)$. Figure \ref{Fig: CaptureVelocity}(c) shows how $v_c$ depends on the detuning, $\Delta_{11}$, when $I_t=260$~mW/cm$^2$ and $B'=39$~G/cm. The capture velocity has a maximum when $\Delta_{11}/(2\pi)\approx 15$~MHz. Across all parameters, the highest capture velocity measured is $v_c=3.8(1)$~m/s. This is an order of magnitude smaller than is typical for a type-I MOT, reflecting the blue-detuned MOT's exclusive reliance on polarization-gradient forces. Nevertheless, over the whole range of the parameters explored, the capture velocity is at least 10 times higher than the mean velocity in the blue-detuned MOT~\cite{Jarvis2018}, so the MOT is stable against the evaporative loss of atoms. 

\subsection{Cold collisions in the blue-detuned MOT}
After loading the blue-detuned MOT, the number of trapped atoms, $N$, decays. We model this using the rate equation,
\begin{equation}
\frac{dN}{dt}=-\gamma N - \beta \int n(\vec{r})^2d^3r,
\end{equation}
where $\gamma$ is the one-body loss rate due to collisions with untrapped atoms, and $n(\vec{r})$ is the atom number density at a position $\vec{r}$ in the MOT. The integral is evaluated over the volume containing the trapped atoms. 
\begin{figure}[t]
	\centering
	\includegraphics{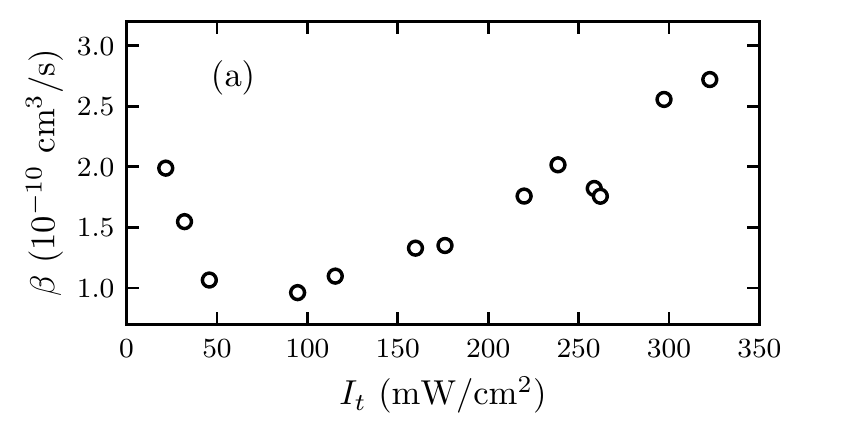}
    \includegraphics{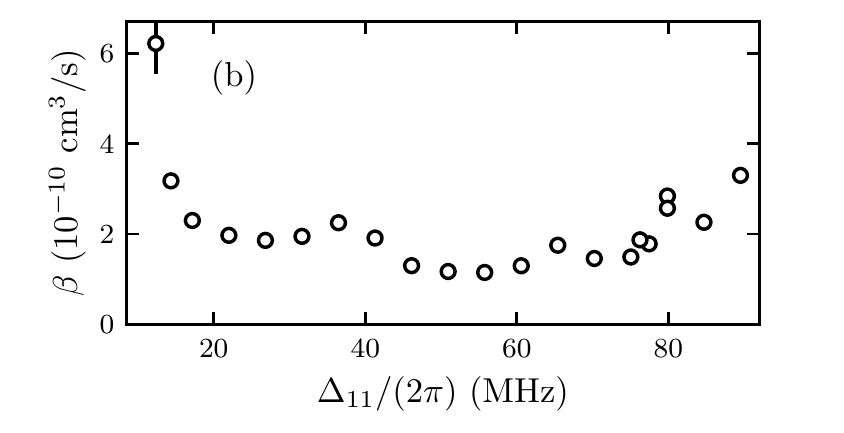}
	\caption{Density-dependent loss rate coefficient, $\beta$ as a function of (a) trap laser intensity, $I_t$, and (b) detuning, $\Delta_{11}$. In (a) the other parameters are $\Delta_{22}/(2\pi) = +12.5$~MHz, $\Delta_{11}/(2\pi)=+26$~MHz, and $B'=87$~G/cm. In (b) the other parameters are $I_t=215$~mW/cm$^2$, $\Delta_{22}/(2\pi)=+11.5$~MHz, and $B'=87$~G/cm.}
	\label{Fig: BlueMOTBetaVsIntensity}
\end{figure}
To measure the loss-rate coefficient $\beta$, the atom density distribution and number of atoms are measured at various times by absorption imaging, as described in \cite{Jarvis2018}. There, we measured $\beta=1.75(35)\times 10^{-10}$~cm$^3$s$^{-1}$ when $I_{t}=240$~mW/cm$^{2}$, $\Delta_{11}/(2\pi)=26$~MHz, $\Delta_{22}/(2\pi) = 12$~MHz and $B'=48$~G/cm. Here, we present a more extensive study of how $\beta$ depends on intensity and detuning. 

Figure \ref{Fig: BlueMOTBetaVsIntensity}(a) shows how $\beta$ depends on $I_t$. As $I_t$ is lowered from a high value, $\beta$ decreases in proportion, reaching a minimum of about $1.0\times 10^{-10}$~cm$^3$s$^{-1}$ when $I_t$ is between 50 and 100~mW/cm$^2$. At lower intensities, $\beta$ increases again, suggesting the onset of an extra loss mechanism. This dependence on $I_{t}$ is qualitatively similar to that measured for type-I MOTs~\cite{Wallace1992,Gensemer1997}, but the variation we observe with $I_{t}$ is far smaller, and our values of $\beta$ are 10-100 times larger for all but the lowest of intensities used in the type-I MOTs. The loss rate coefficient for a type-I MOT operated at $\Delta\simeq\Gamma$ typically varies with intensity by two orders of magnitude~\cite{Wallace1992}, with the highest $\beta$ measured at the lowest intensities and attributed to hyperfine-state changing collisions between ground-state atoms \cite{Shang1994, Ritchie1995, Williamson1995} that release enough energy to eject the colliding atoms from the trap. A collision where only one atom changes hyperfine state imparts a kinetic energy of $\frac{1}{2}mv^2=\Delta E_{\textrm{hfs}}/2$, where $\Delta E_{\textrm{hfs}}=6.8$~GHz is the ground-state hyperfine splitting, giving $v=5.6$~m/s. This is small compared to the capture velocity of a type-I MOT, except for low intensities, but exceeds the capture velocity we have measured in the blue-detuned MOT for all intensities (see Fig.~\ref{Fig: CaptureVelocity}). Therefore, the capture velocity of the blue-detuned MOT is insufficient to recapture the products of any inelastic collision between ground-state atoms, which probably explains the high value of $\beta$ we measure for all $I_{t}$, and the relatively small variation of $\beta$ with $I_{t}$. The additional loss mechanism we observe at low $I_{t}$ could be due to Zeeman-state-changing collisions in the wings of the distribution where the magnetic field is highest, or hyperfine-state-changing collisions involving excited-state atoms. The increasing value of $\beta$ with $I_{t}>50$~mW/cm$^{2}$ may be due to the increased rate of collisions between atoms confined at a higher temperature, since $\beta=\langle \sigma(v)v\rangle$. We have previously measured a linear dependence of the temperature of the blue-detuned MOT on $I_{t}$ over this range~\cite{Jarvis2018}. 

Figure \ref{Fig: BlueMOTBetaVsIntensity}(b) shows how $\beta$ depends on the detuning $\Delta_{11}$. As this detuning is increased, $\beta$ is generally observed to fall, reaching its lowest value at $\Delta_{11}/(2\pi)\approx 55$~MHz. Interestingly, the capture velocity is also observed to decrease over a similar range of detuning, falling to approximately half its maximum value, as can be seen from Fig.~\ref{Fig: CaptureVelocity}(c). We might expect the loss rate to increase as the trap gets shallower, so the decreasing loss rate over this range of detuning cannot be a result of the changing trap depth. Instead, the decrease in $\beta$ may be due to a light-assisted process becoming less efficient or an optical shielding process becoming more efficient. Similar measurements have previously been made for type-II MOTs of Na and $^{85}$Rb. For a red-detuned type-II MOT of $^{85}$Rb, a value of $\beta=9.1(7)\times 10^{-9}$~cm$^3$~s$^{-1}$ has been reported \cite{Tiwari2008}, some 90 times larger than the minimum value we measure. The difference is partly attributable to the difference in temperatures between the two MOTs. Here, the temperature is about 30 times smaller than reported in \cite{Tiwari2008}, so atoms at a given density collide less frequently.

\section{Red-detuned MOTs}
Now we present observations of a number of new trapping configurations that use red-detuned light. As shown in Fig.~\ref{Fig: TrappingConfigurations}, these MOTs are grouped into pairs according to which excited-state hyperfine level $F'$ the ground-state hyperfine level $F=1$ is coupled to.

\subsection{Red MOT A}
\begin{figure}[t]
	\centering
	\includegraphics{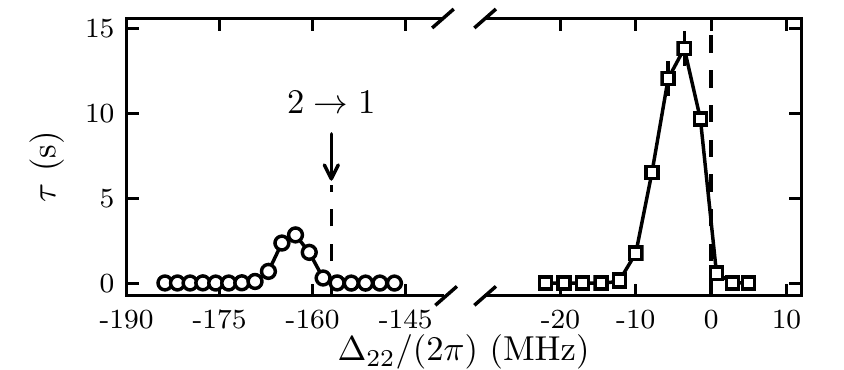}
	\caption{The lifetime of atoms loaded into red MOT A as a function of $\Delta_{22}$. The frequency of the transition to $F'=1$ is indicated. Two sets of data are shown. For each set ${\cal L}_1$ is tuned to maximise the lifetime of the MOT. For the circular markers this optimised detuning is $\Delta_{12}/(2\pi)=-26$~MHz and for the square markers the detuning is $\Delta_{12}/(2\pi)=-36$~MHz. In both cases the total trap intensity is $I_t=260$~mW/cm$^2$ and the field-gradient is $B'=44$~G/cm.}
	\label{Fig: MOTALifetimeVsDetuning}
\end{figure}

\begin{figure}[t]
	\centering
	\includegraphics{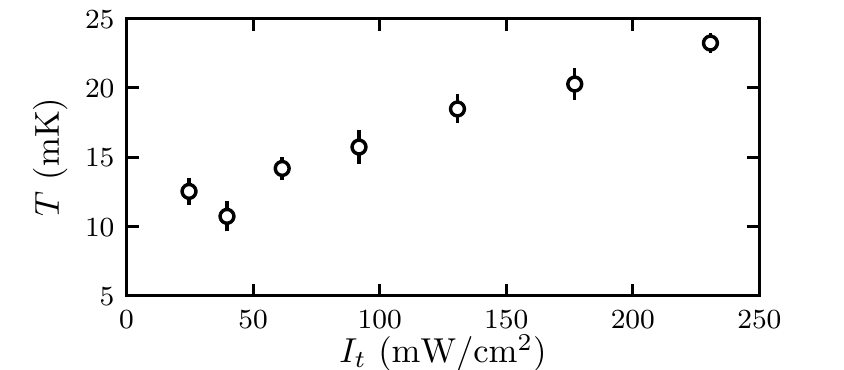}
	\caption{The temperature of atoms confined in MOT A as a function of trap laser intensity. The other trap parameters are $\Delta_{12}/(2\pi)=-27$~MHz, $\Delta_{22}/(2\pi)=-4$~MHz and $B'=44$~G/cm.}
	\label{Fig: MOTATemperatureVsIntensity}
\end{figure}

Red MOT A has a hybrid type-I/type-II nature. The restoring beams are polarised to drive $\Delta m_F=-1$ transitions, the same as for the type-I MOT. The $F=1$ ground state is coupled to $F'=2$, so there are no dark ground-states in this hyperfine level. Consequently, Doppler-cooling can proceed more efficiently than in the pure type-II configurations. ${\cal L}_2$ can be red-detuned from $F'=2$ or red-detuned from $F'=1$, both of which give a trap lifetime of many seconds, as shown in Fig.~\ref{Fig: MOTALifetimeVsDetuning}. The lifetime is longest, up to 14~s, when $\Delta_{22}/(2\pi)\approx-5$~MHz In this configuration, the MOT captures atoms from the background vapour, though the steady-state trapped atom number is much lower than for the type-I MOT. If a larger number of atoms is transferred from the type-I MOT into red MOT A, the number of atoms decays towards the steady-state value. This initial decay is used to measure a trap lifetime. When ${\cal L}_2$ is tuned close to $F'=2$, both ground states are coupled to the same excited state, which decays with equal probability to $F=1$ and $F=2$. The ground-state populations therefore tend to equalize. Because the detuning is negative, atoms in $F=2$ are heated by polarisation-gradient forces. Fig.~\ref{Fig: MOTATemperatureVsIntensity} shows the temperature of red MOT A as a function of $I_t$. The temperature is measured from a sequence of absorption images recorded at fixed intervals during a ballistic expansion, using Gaussian fits to the density distribution recorded in each image. The whole sequence is repeated five times. The temperatures measured here are orders of magnitude higher than those observed in the type-I MOT or blue-detuned type-II MOTs, but comparable to those observed before in red-detuned type-II atomic \cite{Tiwari2008} and molecular MOTs \cite{Williams2017}. 
\subsection{Red MOT B}
\begin{figure}[t]
	\centering
	\includegraphics{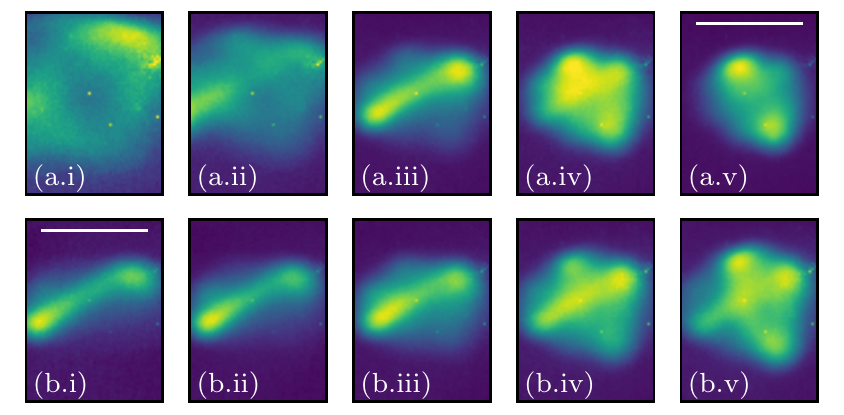}
	\caption{Fluorescence images of atoms in MOT B. In the top row the magnetic-field gradient is increased from left to right. The field gradients used in (a.i)-(a.v) are $B'=28, 43, 50, 58$ and 65~G/cm, respectively. The intensity is $I_t=103$~mW/cm$^2$. In the bottom row the trap intensity is increased from left to right. The values in (b.i)-(b.v) are $I_t = 28, 49, 82, 128,$ and $165$~mW/cm$^2$. The field gradient is $B'=53$~G/cm. In all cases $\Delta_{11}/(2\pi)=-25$~MHz, and $\Delta_{22}/(2\pi)=-12$~MHz. The solid white lines in (a.v) and (b.i) are scale bars of length 5~mm.}
	\label{Fig: MOTBFluorescenceImages}
\end{figure}

Now we discuss red MOT B, which is the red-detuned counterpart to the blue-detuned MOT. Here, the lasers are red-detuned from $F=1\to F'=1$ and $F=2\to F'=2$. The polarisation of the MOT light is reversed relative to that used for a blue detuning in order to preserve a confining force. There is Doppler cooling, but atoms in both ground states are heated by polarisation-gradient forces. We find that this MOT requires a large magnetic-field gradient and high laser intensity to be stable.  Figure \ref{Fig: MOTBFluorescenceImages} shows a series of images showing atoms confined in the MOT as the intensity and field-gradient are varied. The image plane is parallel to the axis of the MOT and is aligned at 45$^\circ$ to the radial beams. Fluorescence from the MOT is imaged onto a CCD camera during a 5~ms exposure. The cloud of atoms is very large in extent compared to the trapping configurations discussed so far, and striking ring-like structures are formed. Notably, atoms are always absent from the centre of the MOT. The top row of images shows that increasing the field gradient causes the diameter of the ring-structure to decrease and its orientation to change. The bottom row shows that increasing the intensity eventually causes a bifurcation into a double-ring structure. Similar behavior has been observed before in type-I MOTs, where a deliberate misalignment of the MOT beams can induce transitions between static and dynamic behavior. Both continuous ring structures and clumped orbital modes have been observed \cite{Felinto1999,Walker1990}, both of which differ from the structures observed here by the presence of a central clump of atoms. This behaviour in the type-I MOT was attributed to multiple photon scattering in the cloud. In the present investigation, the same MOT beams are used as in the other trapping configurations, where the density distributions are roughly Gaussian, so any misalignment must be small. However, since the temperature of the atoms in this MOT is so high, the trap volume explored by the atoms is much larger than typical. At the outermost regions of the trap a net torque is more readily imparted to the atoms by any small misalignment of the beams. 
We note that in red-detuned type-II MOTs, atoms are damped towards a special, non-zero velocity where the Doppler and polarization-gradient forces are in equilibrium~\cite{Devlin2016}. One way to maintain this special speed is to move in circular trajectories, orbiting the trap centre. We have performed trajectory simulations using the position-dependent and velocity-dependent force profiles calculated by solving the optical Bloch equations for this system, as described in \cite{Jarvis2018}. These simulations show that an initially random distribution of atoms quickly forms itself into a ring with atoms circulating about the axis of strongest confinement, supporting our explanation for the structure we observe. The bifurcation to a double-ring structure could be due to the radiation pressure between atoms, which is not captured by our simple model. 

\begin{figure}[t]
	\centering
	\includegraphics{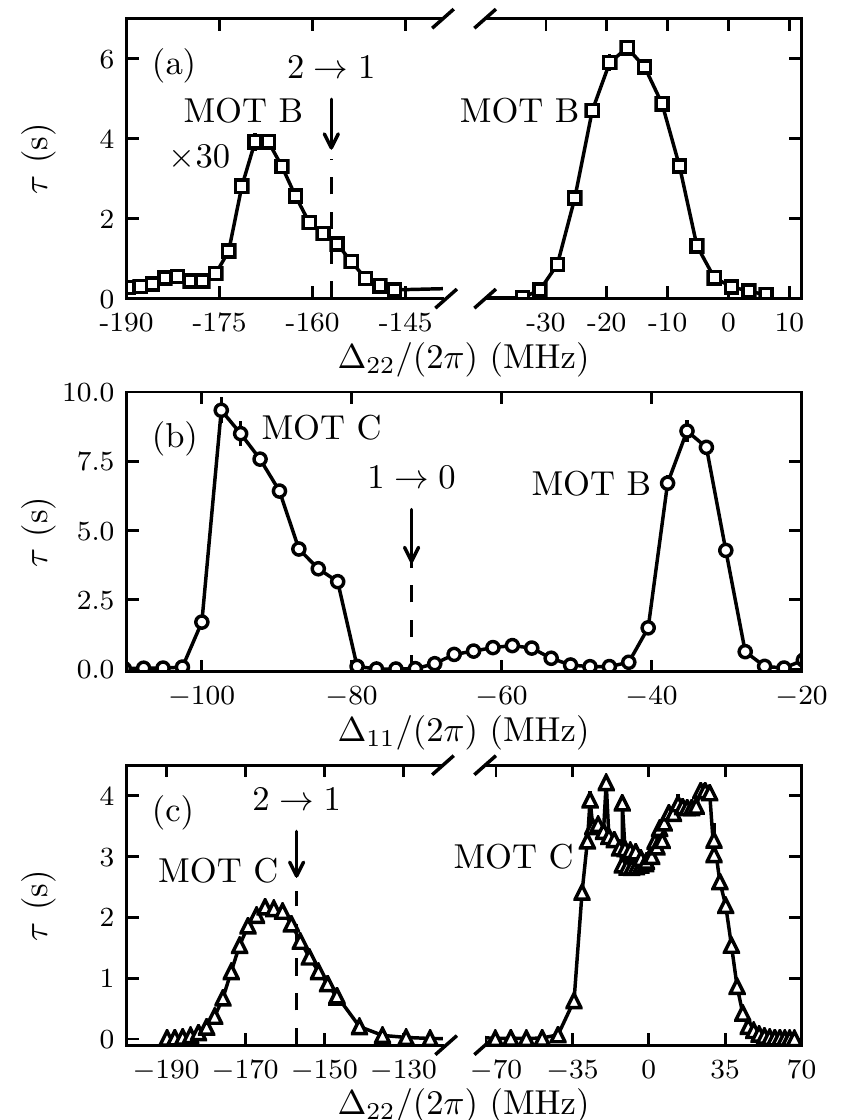}
	\caption{The lifetimes of atoms loaded into MOTs B and C as a function of laser detunings. In (a) ${\cal L}_2$ is scanned as ${\cal L}_1$ is held fixed at $\Delta_{11}/(2\pi)=-31$~MHz. Then, in (b), ${\cal L}_2$ is fixed at $\Delta_{22}/(2\pi)=-17$~MHz and ${\cal L}_1$ is scanned. MOT B appears when $-40\leq\Delta_{11}/(2\pi)\leq-30$~MHz. As $\Delta_{11}$ is made more negative, two more regions of stable trapping appear that are red-detuned and blue-detuned relative to the $F=1\to F'=0$ transition. These are MOT C. A maximum lifetime for MOT C is obtained when $\Delta_{11}/(2\pi)=-97$~MHz, which is equivalent to $\Delta_{10}/(2\pi)=-25$~MHz. Finally, with ${\cal L}_1$ fixed at $\Delta_{11}/(2\pi)=-91$~MHz, ${\cal L}_2$ is again scanned to obtain the data in (c).}
	\label{Fig: ScanningMOTs}
\end{figure}

We turn now to the lifetime of MOT B. For $B'<50$~G/cm, trapped atoms form a diffuse cloud with a short lifetime. As $B'$ is increased from this value, the confinement increases and the lifetime rises quickly, reaching a maximum of about 5~s at a field gradient of 65~G/cm. At even higher $B'$ the lifetime again begins to fall, possibly due to intra-trap collisions.
Figure \ref{Fig: ScanningMOTs} shows the MOT lifetime as the detuning of each of the lasers is scanned in turn. Figure \ref{Fig: ScanningMOTs}(a) shows the lifetime as a function of $\Delta_{22}$, with $\Delta_{11}/(2\pi)=-31$~MHz. We find that ${\cal L}_2$ can be red-detuned from either $F'=1$ or $F'=2$, but the lifetime is many times higher in the latter case, acquiring a maximum value of $\tau\approx6$~s. Figure \ref{Fig: ScanningMOTs}(b) shows similar data as a function of $\Delta_{11}$, when $\Delta_{22}/(2\pi)=-17$~MHz. The lifetime of MOT B is largest when $-40\leq\Delta_{11}/(2\pi)\leq-30$~MHz. We observe no MOT when $\Delta_{11}\approx-50$~MHz, but the MOT reappears as ${\cal L}_1$ is tuned towards $F'=0$. In this region of detuning, the maximum lifetime is  about 1~s, falling back to zero when the laser is resonant with $F'=0$. When ${\cal L}_1$ is red-detuned from $F'=0$, the lifetime quickly increases to  high values and this MOT is referred to as red MOT C.

\subsection{Red MOT C}

Red MOT C is the final trapping configuration discussed. This trap is obtained when ${\cal L}_1$ is tuned close to $F'=0$ and ${\cal L}_2$ is tuned close to either $F'=2$ or $F'=1$, with the former giving longer lifetimes. There should not be any confining force when the excited state has no Zeeman splitting~\cite{Tarbutt2015}, so we would not expect ${\cal L}_1$ to contribute any confinement. Moreover, when ${\cal L}_2$ is tuned near $F'=2$, we find that the lifetime is insensitive to the sign of $\Delta_{22}$, implying that ${\cal L}_2$ is not responsible for the confinement either.  A similar observation has previously been reported for sodium \cite{Nasyrov2001}, where the confinement was attributed to an alignment of the orientation of the ground-state atoms caused by a magnetic-field induced asymmetry in the decay of atoms from $F'=1$. The effect is the result of the mixing of excited-state hyperfine levels by the magnetic field. The force due to the magnetic-field-induced mixing is predicted to be proportional to the third-power of the magnetic-field strength, and inversely proportional to the hyperfine splitting. In $^{87}$Rb, the hyperfine spacing between $F'=0$ and $F'=1$ is 87~MHz, whereas in sodium the equivalent spacing is only 16~MHz. This could explain why MOT C is diffuse, being comparable in extent to the fluorescence images of MOT B shown in Fig. \ref{Fig: MOTBFluorescenceImages}, despite the large field gradients used.

\section{Summary and suggestions}

For each of the alkali metal atoms there is just a single type-I MOT, and these have been studied extensively. There exist many more type-II trapping configurations, few of which have been reported previously. Here we have investigated a number of these configurations that have not been observed before, and have presented their properties which can vary enormously and sometimes be quite extreme. The configurations studied include both blue-detuned and red-detuned type-II MOTs. The latter are diffuse and hot, with temperatures a thousand times higher than for the blue-detuned configurations, which have the favorable properties of low temperatures and high densities. This work consolidates the recent progress made in understanding type-II MOTs, and verifies predictions about the conditions required for these MOTs~\cite{Tarbutt2015,Devlin2016}.

The various configurations reported have different distributions of populations among the ground-state sub-levels and could be used to investigate collisions between trapped atoms in a variety of new settings. For example, it has previously been demonstrated that collisions between ground-state atoms can be optically shielded by blue-detuned light \cite{Sanchez-Villicana1995}. Conversely, red-detuned light can excite ground-state atoms to an attractive molecular potential, where they accelerate towards each other, approach at short distances, and may undergo an inelastic process leading to trap loss. In the D2 line the excited state hyperfine intervals are small enough that colliding atoms in a blue-detuned MOT can still be excited to an attractive molecular potential of a higher-lying hyperfine state, and undergo inelastic collisions that eject them from the trap. However, a previous investigation has shown that trap loss due to light-assisted collisions is reduced by over an order of magnitude from its near-resonant value for a detuning $|\Delta|>800$~MHz, which is about the size of the D1-line hyperfine interval in $^{87}$Rb. Therefore, light-assisted collisions might be strongly suppressed in such a D1 MOT, while the near-resonant blue-detuned light could optically shield inelastic collisions between ground-state atoms. Such a MOT could be formed using a similar procedure to that demonstrated here. The blue-detuned MOT could also be applied to molecules, where type-II transitions are always used. If that works, impressive increases in the phase-space density of molecular MOTs might be possible.  

Underlying data may be accessed from Zenodo~\footnote{\lowercase{10.5281/zenodo.1317582}} and used under the Creative Commons CCZero license.

\acknowledgements
This research has received funding
from EPSRC under Grants No. EP/M027716/1 and No. EP/
P01058X/1.

\bibliography{MOTMenagerieBibliography}
\end{document}